\documentstyle[12pt]{article}
\begin{document}

\thispagestyle{empty}
\rightline{NSF-ITP-94-29}
\rightline{hep-th/9404008}

\vskip 3.5 cm

\begin{center}
{\large\bf  FREE FERMION REPRESENTATION OF A \break
BOUNDARY CONFORMAL FIELD THEORY} \break

\vskip 1.5 cm
{\bf Joseph Polchinski} \footnote{Electronic address:
joep@sbitp.itp.ucsb.edu}
and {\bf L{\'a}rus Thorlacius} \footnote{Electronic address:
larus@sbitp.itp.ucsb.edu}

\vskip 0.3 cm
{\sl Institute for Theoretical Physics  \break
University of California  \break
Santa Barbara, CA 93106}  \break

\end{center}
\vskip 0.6 cm

\begin{quote}
The theory of a massless two-dimensional scalar field with a
periodic boundary interaction is considered.  At a critical
value of the period this system defines a conformal field
theory and can be re-expressed in terms of free fermions,
which provide a simple realization of a hidden $SU(2)$
symmetry of the original theory.  The partition function and
the boundary $S$-matrix can be computed exactly as a function
of the strength of the boundary interaction.
We first consider open strings with one interacting and one
Dirichlet boundary, and then with two interacting boundaries.
The latter corresponds to motion in a periodic tachyon
background, and the spectrum exhibits an interesting band
structure which interpolates between free propagation and
tight binding as the interaction strength is varied.
\end{quote} \vskip1cm

\normalsize

\newpage

\section{Introduction}

Boundary conformal field theories \cite{cardy} find applications in
different branches of physics ranging from condensed matter
systems \cite{calleg}-\cite{kanfis} to particle physics
\cite{callan}, \cite{afsa} and string theory \cite{clny},\cite{caltho}.
A simple but interesting example is provided by a scalar field
in two spacetime dimensions which is free except for a periodic
interaction at a boundary.  This model arises in the
Caldeira-Leggett \cite{calleg} description of the dissipative quantum
mechanics of a particle which is moving in a periodic potential.
In this context, a perturbative renormalization group analysis,
carried out by Fisher and Zwerger \cite{fiszwe} and by Guinea
{\it et al.} \cite{ghm}, showed that there is a critical value of the
potential period where the system undergoes a localization transition.
The same model comes up in the study of open string theory in
background fields \cite{caltho}.  Here the critical potential
corresponds to a particular solution of the equations of motion
of open string field theory.  From the string theory point of
view it is natural to expect the critical theory to exhibit not
only scale invariance but in fact an infinite dimensional
symmetry generated by arbitrary reparametrizations of the
boundary.

In a recent paper this problem was revisited by Callan and
Klebanov \cite{calkle} who phrased it in the language of
two-dimensional conformal field theory with boundary interactions.
They obtained partial results for the partition function and
scattering amplitudes for a boundary potential with the critical
period and made a strong case for the exact conformal invariance
of the model.
In the present paper we analyze the model further and show
how it can be rewritten as a theory of free fermions.  The
periodic potential translates into a fermion mass term,
localized at the boundary, which twists the fermion boundary
conditions by a coupling dependent phase.  The exact partition
function is easily calculated in the fermion language and
scattering amplitudes can be obtained in a straightforward
manner.  The fermions form a representation of a level one
$SU(2)$ current algebra.  This symmetry is not immediately
apparent in the original boson theory although it was clearly
emerging in the work of Callan and Klebanov \cite{calkle}.

In sections~2 and~3 we consider open strings with one interacting and one
Dirichlet boundary.  In section~4 we briefly discuss the boundary $S$-matrix.
In section~5 we consider the case of two interacting boundaries and
find the band structure in the spectrum.  Various subtleties, including
cocycles, the precise mapping between bosonic and fermionic Hilbert spaces,
and operator ordering, arise and are dealt with.
After obtaining many of these results we received a paper by Callan
{\it et al.} \cite{caletal} who have also obtained an exact solution of
the theory by using the underlying $SU(2)$ algebra.

\section{The system}

We are interested in the physics of a massless scalar field
living on a two-dimensional spacetime with a boundary, where
it is subject to a periodic potential.  The Lagrangian, in units
where $\alpha' = 2$, is:
\begin{equation}\label{eqone}
L = {1\over 8\pi} \int_0^l d\sigma \, (\partial_\mu X)^2
-{1\over 2} \Bigl(g e^{iX(0)/\sqrt{2}}
+ \bar g e^{-iX(0)/\sqrt{2}} \Bigr)\,,
\end{equation}
where $g$ is a complex parameter which dials the strength of
the boundary interaction.  Following Callan and Klebanov \cite{calkle}
we include a second boundary at $\sigma =l$ and impose Dirichlet
boundary conditions there to control infrared behavior.
The period in (\ref{eqone}) is chosen such that the potential has dimension
one under boundary scaling and thus defines a marginal perturbation
on the free theory \cite{fiszwe},\cite{ghm}.

We find it convenient to map the theory (\ref{eqone}), which is defined
on a strip of width $l$ into a chiral theory which lives on circle
of circumference $2l$.  To see how this comes about let us first
consider the $g=0$ theory where the boundary interaction has been
turned off and the scalar field satisfies a Neumann condition at
one boundary and a Dirichlet condition at the other.  Away from
the boundaries a free field can be written as a sum of left- and
right-moving components: $X(\sigma,t)=X_L(\sigma+t)+X_R(\sigma-t)$.
The Neumann boundary condition at $\sigma=0$
determines the right-movers in terms of the left-movers,
$$ X_R( 0+t) = X_L( 0- t).$$
Thus we can work with left-movers on the circle.
The Dirichlet condition on $X(\sigma, t)$ at $\sigma = l$ then implies that
$$ X_R(l - t) = - X_L(l + t),$$ with
opposite sign from the Neumann condition.
The combination of the two boundary conditions implies that
$$ X_L(\sigma + 2l) = - X_L(\sigma),$$ so the left-mover is antiperiodic
on the circle of circumference $2l$.

The boundary interaction is easily expressed in terms of the
left-moving field alone,
\begin{equation}\label{eqtwo}
L_{g} = -{1\over 2} \Bigl(g e^{i\sqrt{2}X_L(\sigma^+)}
       \ +\ \bar g e^{-i\sqrt{2}X_L(\sigma^+)} \Bigr)
\Big\vert_{\sigma=0}\, .
\end{equation}
Note that
because $X(0,t)= 2X_L(\sigma^+)\vert_{\sigma=0}$
the coefficient in the exponential has changed.\footnote
{More generally, $e^{i k X(0,t)} = e^{i k_L X_L(t)}$ with $k_L = 2k$.
Thus we must, and will, be careful to distinguish $k_L$ and $k$.}
The operators $e^{\pm i\sqrt{2}X_L(\sigma^+)}$ have conformal
dimension $(1,0)$ and along with
${i\over \sqrt{2}}\partial_+X_L(\sigma^+)$ they form
the currents of a left-moving $SU(2)$ algebra.  It is this
symmetry which enables the exact solution of the model.

\section{Exact solution in terms of fermions}

We now show how to express the interaction in terms of fermions.
Since the scaling dimension of the boundary potential is $(1,0)$
we will need two independent left-moving fermions, each of
dimension $({1\over 2},0)$.  The first step is to introduce an
auxilliary anti-periodic left-moving boson $Y_L(\sigma^+)$.
We then define a pair $\Psi$ of left-moving fermions:
\begin{equation}\label{eqthree}
\psi_1\ \sim\ e^{i(Y_L-X_L)/\sqrt{2}}\ \equiv e^{i\phi_{L1}} \,,\qquad
\psi_2\ \sim\ e^{i(Y_L+X_L)/\sqrt{2}}\ \equiv e^{i\phi_{L2}} \,.
\end{equation}
The extra boson $Y_L$ does not appear in the interaction and so decouples,
but is needed for the fermionic representation.\footnote{The representation
of the boundary interaction in terms of free fermions is due to
Guinea {\it et al.} \cite{ghm}.  The extra boson was implicit in their work,
while for our purposes (the partition function and spectrum) we need to
develop the fermi-bose equivalence in much more detail.}
Actually, the
fermionization~(\ref{eqthree}) is not quite right because $\psi_1$ and
$\psi_2$ commute, being constructed from orthogonal linear combinations.
A cocycle is needed.  In familiar examples (such as the Neumann-Neumann
case to be considered in Section \ref{secfive}), this is constructed from
the bosonic zero modes, but here the bosons have no zero modes.
We instead add a two-state system $S$ to the bosonic theory, and then
\begin{equation}
\label{eqthreetrue}
\psi_1\ =\ \mbox{\boldmath $\sigma$}^1 e^{i\phi_{L1}} \,,\qquad
\psi_2\ =\ \mbox{\boldmath $\sigma$}^2 e^{i\phi_{L2}} \,,
\end{equation}
where the cocycle is written in terms of the $\sigma$-matrices acting in $S$.
Boldface is used to distinguish these operators in the space $S$
from ordinary $SU(2)$ matrices as will appear below.

The bosonic theory plus $S$ is
equivalent to the fermionic, as we now verify by comparing partition
functions. The anti-periodic boundary condition on the bosons translates
into conjugation on the fermions:
\begin{equation}\label{eqfour}
\psi_i(\sigma^+ + 2l) =
\psi^\dagger_i(\sigma^+).
\end{equation}
If we work instead with real fermions $\chi_i$, where
$\sqrt{2} \psi_1=\chi_1+i \chi_2$ and $\sqrt{2} \psi_2=\chi_3+i\chi_4$,
then these boundary conditions dictate that we have two periodic
fermions and two anti-periodic fermions in the free theory.
Writing $q=e^{-\pi\beta /l}$, the corresponding free partition function is:
\begin{eqnarray}\label{eqfive}
Z(q) &=& 2\,  q^{1\over 24}\prod_{n=1}^\infty
(1+q^n)^2(1+q^{n-{1\over 2}})^2 \nonumber \\
&=& 2\,  q^{1\over 24}\prod_{n=1}^\infty
(1-q^{n-{1\over 2}})^{-2} \> ,
\end{eqnarray}
equalling that of two antiperiodic bosons plus $S$.
The factor of two in the first line of (\ref{eqfive})
is from the fermionic zero modes, and the one in the
second line is from $S$.
% The generators of a left-moving $SU(2)$ current algebra at level one are
% constructed from the fermions as
% follows: \begin{equation}
% j^a = \Psi^\dagger \sigma^a \Psi \>.
% \end{equation}

The fermionic action
\begin{equation}\label{eqsix}
L_F = \frac{i}{2\pi} \int_{-l}^{l} d\sigma\>
\Psi^\dagger \bigl(\partial_t - \partial_\sigma
 - i {\bf M} \delta(\sigma) \bigr) \Psi \>,
\end{equation}
with $\Psi = \left[ \begin{array}{c} \psi_1 \\ \psi_2 \end{array} \right]$,
$ {\bf M} = \pi( g_1 \sigma^2 + g_2 \sigma^1)$ and $g = g_1 + i g_2$,
is then equivalent to the $X_L$-$Y_L$-$S$ system with interaction
\begin{equation}\label{eqtwonew}
L'_{g} = -\frac{\mbox{\boldmath $\sigma$}^3}{2} \Bigl(g
e^{i\sqrt{2}X_L(\sigma^+)}
       \ +\ \bar g e^{-i\sqrt{2}X_L(\sigma^+)} \Bigr)
\Big\vert_{\sigma=0}\, .
\end{equation}
We thus have a sum of two copies of the interacting theory, with opposite
signs for $g$.  Since the fermions~(\ref{eqthreetrue}) flip the sign of
$\mbox{\boldmath $\sigma$}^3$, the desired single copy can be obtained by
projection on fermion number mod~2.

The normal modes of frequency $\nu$,
\begin{equation}\label{eqeight}
\Psi_\nu(t,\sigma) =
e^{-i\nu t} \tilde\Psi_\nu(\sigma)
+ e^{i\nu t} \tilde\Psi_{-\nu}(\sigma)
\> ,\end{equation}
satisfy
\begin{equation}\label{eqseven}
\tilde\Psi'_\nu =
-i \bigl(\nu +  {\bf M}
\delta(\sigma) \bigr) \tilde\Psi_\nu
\>. \end{equation}
This is ill-defined because the delta-function multiplies $\tilde\Psi$,
which is discontinuous at $\sigma=0$.  This is where the issue of
regularization comes up in the fermion language.
We deal with it by smearing out the delta-function
symmetrically around $\sigma=0$ by a smooth function which
satisfies: $f(\sigma)=f(-\sigma)$ and
$\int_{-l}^l d\sigma\,f(\sigma)=1$.
This prescription produces an unambiguous answer for the
partition function.  In terms of the original bosonic theory, this is
equivalent (in Euclidean time) to extending the interaction analytically in
$\sigma + i\tau$ from $\sigma = 0$ and then smearing the integration
contour.  It is thus the same as the
principle value prescription adopted by Callan {\it et al.}
\cite{caletal}.  The integrated result is then
\begin{equation}
\tilde\Psi_\nu(l) =
e^{-2il\nu - i  {\bf M} } \tilde\Psi_\nu(-l).
\end{equation}
The boundary condition $\tilde\Psi_\nu(l) = \tilde\Psi_{-\nu}^*(-l)$ then
implies
\begin{equation}
\tilde\Psi_\nu(l) =
e^{-4il\nu}e^{ - i  {\bf M} }
e^{i  {\bf M} ^*}
\tilde\Psi_\nu(l) \>,
\end{equation}
thus determining the eigenvalues $\nu$.
A short calculation gives: $e^{4il\nu}= e^{\pm 2i\Delta(g,\bar g)}$,
with
\begin{equation}\label{eqeleven}
\Delta(g,\bar g)= \arcsin{\left[
{g+\bar g\over 2|g|}\,
{\sin{\pi |g| }}
\right]} \>.
\end{equation}
The boundary interaction shifts the frequencies and lifts the
degeneracy of the two pairs of real fermions.

Summing over eigenfrequencies $\nu = (n\pi \pm \Delta)/2l$,
the fermion partition function is
\begin{equation}\label{eqtwelve}
Z^\eta_F(q,\alpha)\ =\ {\rm Tr} [\eta^F q^{H\pi/l}]\ =
\ q^{\alpha^2-{1\over 2}\alpha+{1\over 24}}
\, (1+\eta q^\alpha) \prod_{n=1}^\infty
(1+\eta q^{{n\over 2}+\alpha})
(1+\eta q^{{n\over 2}-\alpha}) \>,
\end{equation}
where the shift $\alpha=\Delta(g,\bar g)/ 2\pi$ is a coupling-dependent
constant.  The required projection which selects out states of odd
fermion number is
\begin{equation}\label{eqthirteen}
Z_F^{odd}={1\over 2} \bigl[Z_F^{(+)}-Z_F^{(-)}\bigr] \>,
\end{equation}
with
$Z_F^{(\pm)}(q,\alpha)=q^{\alpha^2-{1\over 2}\alpha+{1\over 24}}
(1\pm q^\alpha) \prod_{n=1}^\infty
(1\pm q^{{n\over 2}+\alpha})
(1\pm q^{{n\over 2}-\alpha})$.
Changing the sign of the second term in
(\ref{eqthirteen}) selects states of even fermion number
and leads to the same projected partition function
up to an overall sign change of both $g$ and $\bar g$.  The desired
sign~(\ref{eqthirteen}) is most easily determined by comparing with
lowest order perturbation theory.

By using Jacobi's triple product formula, the projected partition
function can be represented as a sum:
\begin{equation}\label{eqfourteen}
Z_F^{odd}(q,\alpha)=
q^{-{1\over 48}}
\prod_{m=1}^\infty (1-q^{m/2})^{-1}
\sum_{n=-\infty}^\infty
q^{{1\over 4}[n+{1\over 2}+2\alpha]^2}
\left[{1+(-1)^n\over 2}\right] \>.
\end{equation}
Finally, to obtain the exact partition function of the original
interacting boson theory we have to divide (\ref{eqfourteen}) by
the partition function,
$Z_0(q)=q^{1\over 48}\prod_{n=1}^\infty
 (1-q^{n-{1\over 2}})^{-1}$,
of the free anti-periodic boson we added to the system:
\begin{equation}\label{eqfifteen}
Z_g(q)={Z_F^{odd}(q,\alpha)\over Z_0(q)}
= q^{-{1\over 24}}
\prod_{m=1}^\infty (1-q^m)^{-1}
\sum_{n=-\infty}^\infty
q^{{1\over 4}[2n+{1\over 2}+
{\Delta(g,\bar g)\over \pi}]^2} \>.
\end{equation}

The terms in the sum can be rearranged to get it into the
form of a sum over Virasoro modules:
\begin{equation}\label{eqsixteen}
Z_g(q) = q^{-{1\over 24}}
\prod_{m=1}^\infty (1-q^m)^{-1}
\sum_{k=0}^\infty
q^{{1\over 4}[k+{1\over 2}+(-1)^k
{\Delta(g,\bar g)\over \pi}]^2} \>.
\end{equation}
This result was obtained independently by Callan
{\it et al.} \cite{caletal} using a different approach.

\section{Boundary $S$-matrix}

It is also straightforward to calculate amplitudes for
scattering off the interacting boundary in this model.
In this case one wants to discuss asymptotic incoming
and outgoing states so we take the Dirichlet boundary
off to $l\rightarrow \infty$.  The map to a chiral
theory is very useful for amplitude calculations and we
also find it convenient to rotate to euclidean signature.
Incident left-movers then have $Re(z)>0$ and the outgoing
states are also left-moving but have $Re(z)<0$.
The boundary interaction is localized on the contour
$Re(z)=0$.  Away from the boundary, left-moving bosons are
created by the insertion of $\partial_zX(z)$
operators into the euclidean path integral.

The boundary interaction with critical period defines a
dimension $(1,0)$ operator in the chiral theory.  It is
integrated along $Re(z)=0$ in the euclidean action but
we are free to deform the contour into the complex
plane.  For the calculation of any given amplitude the
contour can be moved to the right until it surrounds
each of the in-state vertex insertions (or, if preferred,
it can be moved to the left to surround the out-states).
This generates a global $SU(2)$ transformation
$e^{i\pi(gJ_++\bar gJ_-)}$, where
$J_+=\oint {dz\over 2\pi i}\, \psi_1^\dagger \psi_2$ and
$J_-=\oint {dz\over 2\pi i}\, \psi_2^\dagger \psi_1$,
on each in-state.  The effect of the $SU(2)$ rotation on
each state can be computed by using the $SU(2)$ current
algebra, as shown by Callan {\it et al.} \cite{caletal},
or it can be obtained directly from the free fermion
operator product expansion.
Once the effect of the boundary interaction has been captured
by the $SU(2)$ rotations, the amplitude calculation reduces
to an exercise in free field theory.  Since the asymptotic
states themselves are created by $SU(2)$ generators, one can
also utilize the current algebra for this computation.
Callan {\it et al.} \cite{caletal} work out some explicit
examples using this procedure and we will not repeat those
calculations here.
As these authors point out, the $SU(2)$ rotation generates the
soliton operators $e^{i\sqrt{2}X(z)}$ and $e^{-i\sqrt{2}X(z)}$.
These operators create non-perturbative kink states,
where the $X$-field shifts between adjacent minima of
the boundary potential, and amplitudes involving such states
must be included in order to obtain a unitary $S$-matrix.
In the fermion language both $\partial_zX(z)$ and the
soliton operators are represented as fermion bilinears and
thus they all enter on equal footing from the start.
Since the fermion theory is manifestly a free theory it is clear
that we have included all states necessary for unitarity and that
soliton-antisoliton states, or states which shift the
boson by more than one period, are not independent objects.

\section{Periodic Tachyon Backgrounds}\label{secfive}

The boundary conformal field theory can be regarded as an open string
tachyon background.  An open string propagating in such a background
then has Neumann conditions with interactions at both ends.
Each endpoint interaction can be written in terms of the left-mover $X_L$,
but we need to be careful about the relative phase of the two interactions.
The open string mode expansion for $X$ is
\begin{equation}\label{eqt0}
X(\sigma,t) = x + \frac{4\pi}{l}pt + i \sum_{m \neq 0} \frac{\alpha_m}{m}
\Bigl( e^{-i m (t + \sigma) \pi/l} + e^{-i m (t-\sigma) \pi/l} \Bigr) \>,
\end{equation}
while
\begin{equation}\label{eqt00}
X_L(\sigma + t) = \frac{x}{2} + \frac{2\pi}{l} p(t + \sigma)
 + i \sum_{m \neq 0} \frac{\alpha_m}{m}
e^{-i m (t + \sigma) \pi/l} \>.
\end{equation}
The zero mode part of $e^{ik X(\sigma,t)}$ is $e^{i k x + 4\pi i k pt/l}$,
where the symmetric
ordering of $x$ and $p$ is appropriate for a tensor in the $\sigma \pm t$
frame, while that
of $e^{2ik X_L(\sigma+t)}$ is
\begin{equation}\label{eqt01}
e^{i k x + 4\pi ik p(t + \sigma)/l} = e^{i k x + 4\pi ik pt/l}
e^{2\pi ik(2p + k) \sigma/l}.
\end{equation}
The nonzero modes are the same in both operators, so for
$k = 1/\sqrt{2}$ we have
\begin{eqnarray}
e^{i \sqrt{2} X_L( t)} &=& e^{i X(0,t) /\sqrt{2}} \nonumber\\
e^{i \sqrt{2} X_L(l+t) } &=& - e^{i X(l,t) /\sqrt{2}}
e^{2\sqrt{2}\pi i p} \>.
\end{eqnarray}
In terms of the left-moving
boson the interaction is then
\begin{equation}\label{eqt1}
L_{\rm int} = -{g\over 2}
\Bigl( e^{i \sqrt{2} X_L(t) } \ -\ e^{i \sqrt{2} X_L(l+t) }
e^{-2\sqrt{2}\pi i p} \Bigr)
\ +\ {\rm h.c.} \>.
\end{equation}

Again the spectrum is easily obtained either from current algebra or the
free fermi representation, and we work with the latter.
The fermionization is
\begin{equation}\label{eqt2}
\psi_1\ =\ e^{i\phi_{L1}} \,,\qquad
\psi_2\ =\ e^{i\phi_{L2}} e^{-2\sqrt{2}\pi i  p_X} \, ,
\end{equation}
where we remind the reader of the notation $p_X = p_{LX}/2 =
(p_{L2} - p_{L1})/2\sqrt{2}$.
Now the cocycle is constructed from the bosonic zero mode and the theories
are equivalent without additional discrete degrees of freedom.\footnote
{There is some freedom in the choice of cocyle, and we have chosen to
construct it entirely from $p_X$ so that $Y$ will remain decoupled.}
However, we must be careful about the correspondence between bosonic momenta
and fermionic boundary conditions.  A complex left-moving fermion with
boundary condition
\begin{equation}
\psi_i(l) = -e^{2\pi i \zeta_i} \psi_i(-l)
\end{equation}
is equivalent to a left-moving boson with momentum $p_{Li} = \zeta_i$
mod ${\bf Z}$.  This is evident from consideration of the vertex operators,
for example $e^{i \zeta \phi_{Li}}$
for the ground state, or from the mode expansion (\ref{eqt00})
for $\phi_{Li}$.  For reasons soon to be explained, it is useful to restrict
further to states of even fermion number relative to the ground state.
This is equivalent to the space of bosonic states with $p_{Li} = \zeta_i$
mod $2{\bf Z}$.  The respective partition functions are
\begin{eqnarray}\label{eqt3}
Z_0(q,\zeta) &=& q^{\frac{1}{2}\zeta^2 - \frac{1}{24}}
\frac{1}{2} \sum_{\pm}
\prod_{n=1}^{\infty} (1 \pm q^{n + \zeta - \frac{1}{2}})
(1 \pm q^{n - \zeta - \frac{1}{2}}) \nonumber\\
&=&
\sum_{m = -\infty}^\infty q^{\frac{1}{2}(\zeta + 2m)^2 - \frac{1}{24}}
\prod_{n=1}^\infty (1-q^n)^{-1}.
\end{eqnarray}
We will focus on the case that the physical boson $X_L$ is noncompact and
take the extra boson $Y_L$ also noncompact, and so must integrate
$(2\pi)^{-2} \int\!\int_0^2 d\zeta_1\, d\zeta_2$ to cover the full space.

In fermionic form the interaction is
\begin{equation}\label{eqt1c}
L'_{\rm int} = -{g\over 2}
\Bigl( \psi_1^\dagger(t) \psi_2(t) e^{2\sqrt{2}\pi i p_X}
\ -\ \psi_1^\dagger(l+t) \psi_2(l+t) \Bigr)
\ +\ {\rm h.c.} \>.
\end{equation}
This is inconvenient because $e^{2\sqrt{2}\pi i p_X}$ anticommutes with the
fermi fields.  However, we are free to consider instead
\begin{equation}\label{eqt2c}
L''_{\rm int} = -{g\over 2}
\Bigl( \psi_1^\dagger(t) \psi_2(t) e^{2\sqrt{2}\pi i p_X} (-1)^F
\ -\ \psi_1^\dagger(l+t) \psi_2(l+t) \Bigr)
\ +\ {\rm h.c.} \>,
\end{equation}
since we have expressed the bosonic space in terms of states of even fermion
number.  The combination $e^{2\sqrt{2}\pi i p_X} (-1)^F = w$ commutes with
the fermi fields, and is simply equal to $e^{i\pi(\zeta_2 - \zeta_1)}$ in
a given sector.

The fermionic action for the interacting theory is now
\begin{equation}\label{eqt2a}
L_{BG,F} = \frac{i}{2\pi} \int_{-l}^{l} d\sigma\>
\Psi^\dagger \bigl(\partial_t - \partial_\sigma
 + i  {\bf N}_1 \delta(\sigma)
 - i  {\bf N}_2 \delta(\sigma - l) \bigr) \Psi \>,
\end{equation}
where
\begin{equation}
 {\bf N}_1 =  \pi \left[ \begin{array}{cc}
0 &  wg \\ \bar w \bar g & 0 \end{array} \right]\>, \qquad
 {\bf N}_2 = \pi \left[ \begin{array}{cc}
0 &  g \\  \bar g & 0 \end{array} \right]\>.
\end{equation}
The normal modes now satisfy the periodicity
\begin{equation}
\tilde\Psi_\nu(l) = -e^{2\pi i (\zeta_+ + \sigma^3 \zeta_- )}
\tilde\Psi_\nu(-l) \>,
\end{equation}
where $\zeta_{\pm} = \frac{1}{2}(\zeta_1 \pm \zeta_2)$.
The equation of motion gives
\begin{equation}
\tilde\Psi_\nu(l) = e^{-2 i l\nu}
e^{-i {\bf N}_2 } e^{i {\bf N}_1 }
\tilde\Psi_\nu(-l) \>.
\end{equation}
Together these give the eigenvalue equation
\begin{equation}
-e^{-2 i l\nu} \tilde\Psi_\nu(-l) =
e^{-i {\bf N}_1} e^{i {\bf N}_2}
e^{2\pi i (\zeta_+ + \sigma^3 \zeta_- )}
\tilde\Psi_\nu(-l) \>.
\end{equation}
The solution is
\begin{equation}\label{lambda}
\frac{l\nu}{\pi} = n + \frac{1}{2} - \zeta_+ \pm \lambda \>,\qquad
\sin \pi \lambda = \cos \pi |g|\, \sin \pi \zeta_-  \>,
\end{equation}
with the value of $\lambda$ determined by continuity from
$\lambda = \zeta_-$ at $g=0$.

The fermionic partition function is then the product
of two copies of~(\ref{eqt3}),
with $\zeta = \zeta_+ \pm \lambda$.  In bosonic form this gives
\begin{equation}
Z = \int \!\!\!\int_0^2 \frac{d\zeta_1}{2\pi}
\frac{d\zeta_2}{2\pi} \sum_{m_1,m_2 \in {\bf Z}}
q^{ (\zeta_+ + 2m_+)^2  + (\lambda +2 m_-)^2 -\frac{1}{12}}
\prod_{n=1}^\infty (1 - q^n)^{-2}
\> ,
\end{equation}
where $m_\pm = (m_1 \pm m_2)/2$.
We must now regroup in order to
separate the contribution of the $Y$ boson.  Note first that
the sum on $(m_1,m_2)$ amounts to summing $(m_+,m_-)$ over all integer
points {\it and} all integer points plus $(\frac{1}{2},\frac{1}{2})$.
The periodicity under $\zeta_1 \to \zeta_1 + 2$ takes the form
$m_\pm \to m_\pm + \frac{1}{2}$ and thus we can use this periodicity
to enlarge the integration range to $0 \leq \zeta_1 \leq 4$ while
restricting $(m_+,m_-)$ to the integer points.  We can then use the
manifest periodicity in $\zeta_+$ and $\zeta_-$ mod $2 \bf Z$ to shift
the integration range to $0 \leq \zeta_+ \leq 2$,
$0 \leq \zeta_- \leq 2$ (this is most easily seen by drawing the two
regions).  Finally, $\zeta_+$ and $m_+$ combine into a single variable
$k_Y/\sqrt{2} = \zeta_+ + 2 m_+$ with range $-\infty$ to $\infty$.
Thus,
\begin{equation}
Z = \int_{-\infty}^\infty \frac{dk_Y}{2\pi} \int_0^2
\frac{d\zeta_-}{\sqrt{2}\pi}
\sum_{m_- = -\infty}^{\infty} q^{ \frac{1}{2} k_Y^2
+ (\lambda + 2 m_-)^2
-\frac{1}{12}}
\prod_{n=1}^\infty (1 - q^n)^{-2} \> .
\end{equation}
The desired partition function is
\begin{equation}
\frac{Z}{Z_Y} = \int_0^2
\frac{d\zeta_-}{\sqrt{2}\pi}
\sum_{m_- = -\infty}^{\infty} q^{  \kappa^2 - \frac{1}{24} }
\prod_{n=1}^\infty (1 - q^n)^{-1} \> ,
\end{equation}
where $\kappa = \lambda + 2 m_-$.

As expected for a particle in a periodic
potential, the spectrum has split up into bands.
As $\zeta_-$ runs through its range, $\lambda$ runs through the values
\begin{eqnarray}
&& 0 \leq \lambda \leq r \qquad 1 - r \leq \lambda \leq
1 + r \qquad 2 - r \leq \lambda \leq 2 \>,\nonumber\\
&& 2r = \Bigl| 1 - 2 F(|g|) \Bigr| \>,
\end{eqnarray}
with $F(|g|)$ denoting the fractional part of $|g|$.  In all, the allowed
values of $\kappa$ consists of bands of width $2r$ centered at every
integer, with gaps of width $1- 2r$ between.
When $|g|$ is any integer, $2r = 1$ and the gaps disappear; this corresponds
to free propagation, the particle not seeing the potential.  When $|g|$ is
integer plus $\frac{1}{2}$, the bands have zero width.
This is the tight-binding limit, with no tunnelling between the minima of
the potential.\footnote{This agrees with the observation~\cite{caletal}
that the latter values correspond to the Dirichlet boundary state.}
Thus we see the physics of strings moving in a tachyonic crystal.
We would expect also to see a `phonon' mode; indeed, at $|\kappa| = 1$,
the wavenumber of the potential, there are two massless (weight one)
states, representing the phase (translation) and the magnitude of the
background.

We can extend readily to the theory compactified at radius
$R = q \sqrt{2}$, where $q$ must be an integer because the potential must
respect the periodicity.  The only effect of compactification of open
strings is to restrict to momenta
\begin{equation}
\zeta_- \ =\ - k_X \sqrt{2} \ \in\ \frac{{\bf Z}}{q}  \>.
\end{equation}
At the self-dual radius $q=1$, $\zeta_-$ is an integer and so
eq. (\ref{lambda}) implies that the spectrum is
unaffected by the potential, as found in ref.~\cite{caletal}.

\section*{Acknowledgements}
We would like to thank Curt Callan, Matthew Fisher, and Igor Klebanov for
discussions.  This work was supported in part by National Science Foundation
grants PHY-89-04035 and PHY-91-16964.  L.~T. thanks the Physics Department,
Victoria University at Wellington for hospitality, during the final stage
of this work.


\begin{thebibliography}{100}

\bibitem{cardy} J. L. Cardy, Nucl. Phys. {\bf 240}
(1984), 514.

\bibitem{calleg} A. O. Caldeira and A. J. Leggett,
Ann. of Phys. {\bf 149} (1983), 374.

\bibitem{fiszwe} M. P. A. Fisher and W. Zwerger,
Phys. Rev. {\bf B32} (1985), 6190.

\bibitem{ghm} F. Guinea, V. Hakim, and A. Muramatsu,
Phys. Rev. Lett. {\bf 54} (1985), 263.

\bibitem{afflud} I. Affleck and A. W. W. Ludwig,
Nucl. Phys. {\bf B352} (1991), 849;
Nucl. Phys. {\bf B360} (1991), 641.

\bibitem{kanfis} C. L. Kane and M. P. A. Fisher,
Phys. Rev. {\bf B46} (1992), 15233.

\bibitem{callan} C. G. Callan, in ``Proceedings
of the Fourth International Conference on Quantum
Mechanics in the Light of New Technology'',
S.~Kurihara ed., Japan Physical Society (1992).

\bibitem{afsa} I. Affleck and J. Sagi,
{\it Monopole Catalyzed Baryon Decay: A Boundary
Conformal Field Theory Approach}, British
Columbia University preprint, UBCTP-93-18,
hepth/9311056, November 1993.

\bibitem{clny} C. G. Callan, C. Lovelace,
C. R. Nappi, and S. A. Yost,
Nucl. Phys. {\bf B308} (1988), 221.

\bibitem{caltho} C. G. Callan and L. Thorlacius,
Nucl. Phys. {\bf B329} (1990), 117.

\bibitem{calkle} C. G. Callan and I. R. Klebanov,
{\it Exact C=1 Boundary Conformal Field Theories},
Princeton University preprint, PUPT-1432,
hepth/9311092, November 1993.

\bibitem{caletal} C. G. Callan, I. R. Klebanov,
A. W. W. Ludwig, and J. M. Maldacena,
{\it Exact Solution of a Boundary Conformal
Field Theory}, Princeton University preprint,
PUPT-1450, hepth/9402113, February 1994.

\end{thebibliography}
\end{document}